\documentclass[journal]{IEEEtran}

\ifCLASSINFOpdf
\else
   \usepackage[dvips]{graphicx}
\fi
\usepackage{url}

\usepackage{graphicx}

\usepackage{amsmath,amsfonts}
\usepackage{bm}
\usepackage{algorithmic}
\usepackage{algorithm}
\usepackage{array}
\usepackage{textcomp}
\usepackage{stfloats}
\usepackage{verbatim}
\usepackage{multirow}
\usepackage{booktabs}
\usepackage{makecell}
\usepackage{xcolor}
\usepackage{hyperref}
\usepackage{subcaption}
\usepackage[numbers, sort&compress]{natbib}
\usepackage{balance}

\makeatletter
\newcommand\footnoteref[1]{\protected@xdef\@thefnmark{\ref{#1}}\@footnotemark}
\makeatother

\begin{document}

\title{HiFi-Glot: High-Fidelity Neural Formant Synthesis \\ with Differentiable Resonant Filters}

\author{
Yicheng Gu, \IEEEmembership{Student Member, IEEE},
Pablo P\'{e}rez Zarazaga,
Chaoren Wang,
Zhizheng Wu, \IEEEmembership{Senior Member, IEEE},
Zofia Malisz,
Gustav Eje Henter,
and Lauri Juvela, \IEEEmembership{Member, IEEE}
\thanks{This work was supported by the Swedish Research Council grant no.\ 2017-02861 ``Multimodal encoding of prosodic prominence in voiced and whispered speech'' and partially supported by the Wallenberg AI, Autonomous Systems and Software Program (WASP) funded by the Knut and Alice Wallenberg Foundation. We acknowledge the computational resources provided by the Aalto Science-IT project. We acknowledge the EuroHPC Joint Undertaking for awarding this project access to the EuroHPC supercomputer LUMI, hosted by CSC (Finland) and the LUMI consortium through a EuroHPC Regular Access call. (Corresponding author: Lauri Juvela.)}
\thanks{Yicheng Gu and Lauri Juvela are with the Acoustic Lab, Department of Information Communications Engineering, Aalto University, Espoo, Finland.}
\thanks{Yicheng Gu, Chaoren Wang, and Zhizheng Wu are with the School of Data Science, The Chinese University of Hong Kong, Shenzhen, China.}
\thanks{Pablo P\'{e}rez Zarazaga, Zofia Malisz, and Gustav Eje Henter are with the Department of Speech, Music and Hearing, KTH Royal Institute of Technology, Stockholm, Sweden.}
}

\maketitle

\begin{abstract}
Formant synthesis aims to generate speech with controllable formant structures, enabling precise control of vocal resonance and phonetic features. However, while existing formant synthesis approaches enable precise formant manipulation, they often yield an impoverished speech signal by failing to capture the complex co-occurring acoustic cues essential for naturalness. To address this issue, this letter presents HiFi-Glot, an end-to-end neural formant synthesis system that achieves both precise formant control and high-fidelity speech synthesis. Specifically, the proposed model adopts a source–filter architecture inspired by classical formant synthesis, where a neural vocoder generates the glottal excitation signal, and differentiable resonant filters model the formants to produce the speech waveform. Experiment results demonstrate that our proposed HiFi-Glot model can generate speech with higher perceptual quality and naturalness while exhibiting a more precise control over formant frequencies, outperforming industry-standard formant manipulation tools such as Praat. Code, checkpoints, and representative audio samples are available at \url{https://www.yichenggu.com/HiFi-Glot/}.
\end{abstract}
\vspace{-2pt}
\vspace{-10pt}

\IEEEpeerreviewmaketitle

\section{Introduction}

The ability to explicitly control speech parameters, such as the fundamental frequency and formant frequencies, is crucial for deep learning applications in speech science. For instance, phoneticians and psycholinguists require synthetic stimuli to experimentally disentangle the perceptual effects of different acoustic speech properties~\cite{phonetician-1, phonetician-2}. Meanwhile, transgender voice therapists must adjust speech materials in terms of pitch, resonance, and vocal weight to provide effective guidance for their patients~\cite{premod-model, premod}. To construct such speech material, experts need to control low-level speech parameters such as formant frequencies, spectral tilt, spectral centroid, and energy independently and accurately. However, while speech technologies have made significant progress~\cite{maskgct, emilia-journal} in recent years, the fine-grained control of such parameters has not been thoroughly explored, leaving a notable research gap~\cite{hifiglot-baseline, formant-2}.

Traditionally, speech modification has utilized DSP-based methods. Techniques such as time-domain pitch-synchronous overlap-add (TD-PSOLA) and its extensions~\cite{psola-1, psola-2} remain common for pitch and timing manipulation but cannot control formants. 
Meanwhile, the source-filter model~\cite{source-filter} presents a methodology for speech generation that mimics the human speech production system. It allows independent analysis of the signal's spectral envelope (e.g., formant analysis) and glottal excitation (e.g., pitch analysis). 
Conventional formant synthesizers based on the source-filter model can achieve parametric control over the speech signal~\cite{formant-synthesizer}. 
However, while formant synthesis performs well in specific tasks, it generally fails to generate natural-sounding speech~\cite{formant-synthesizer-2}. Even with advanced source-filter vocoders like STRAIGHT~\cite{straight}, a naturalness gap remains. Another method for manipulating formants is to apply linear predictive inverse filtering, which modifies the formants in the spectral envelope and uses the original residual excitation signal. Praat~\cite{praat} provides a widely used implementation of this. However, imperfect source-filter separation leaves residual formant information in the source signal, leading to artifacts when manipulating formants.

In contrast, deep learning–based approaches generally adopt neural vocoders to synthesize speech from speech parameters. Neural vocoders are neural networks used to transform a set of acoustic features into a speech waveform. 
The most common acoustic input features for neural vocoders are mel-spectrograms~\cite{WaveNet, HiFiGAN}; while mel-spectrograms can effectively represent speech, they do not offer intuitive or disentangled control over its phonetic characteristics. 
To enable more explicit and interpretable control over speech parameters, neural source–filter models are used. 
Control of the source component has been achieved through harmonic-oscillator excitation models~\cite{nsf} and pitch-synchronous synthesis~\cite{puffin}, enabling explicit control regarding the fundamental frequency. 
Complementary to source modeling, the filter component captures the speech spectral envelope, reflecting the vocal tract's resonant characteristics.
Some approaches incorporate this part using all-pole filters~\cite{gelp}, while others use cepstral representations~\cite{sifigan} to parameterize the spectral envelope.

\begin{figure*}[t!]
\centering
\includegraphics[width=\linewidth]{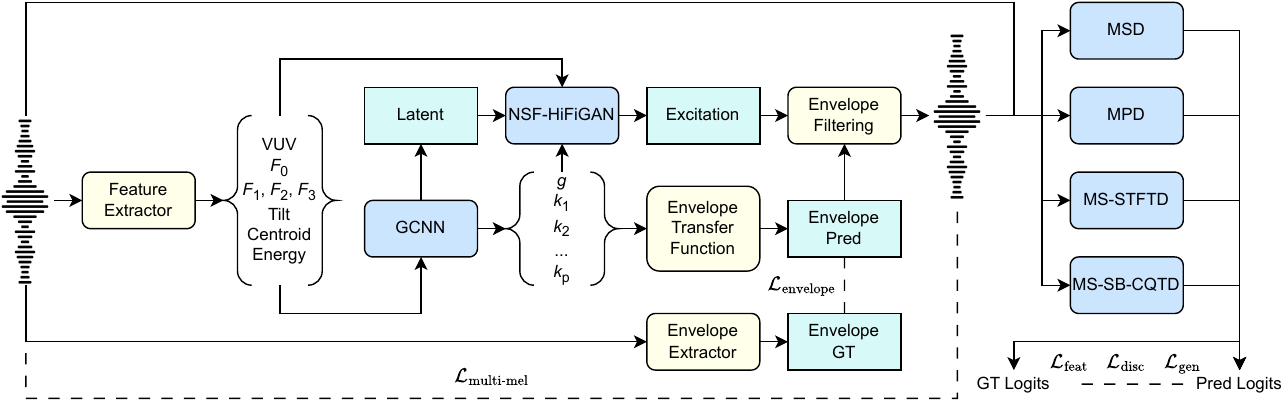}
\caption{
Architecture and training schemes for HiFi-Glot. The model consists of a GCNN feature mapper, an NSF-HiFiGAN decoder, and four different discriminators. Dark-blue blocks represent trainable neural networks, light-blue blocks represent intermediate features, yellow blocks are differentiable DSP modules, and dashed lines indicate loss-function input pairings.
}
\label{fig:model}
\end{figure*}

Building upon these works, several studies have explored neural speech synthesis directly from interpretable speech parameters. In particular, WavebenderGAN~\cite{wavebender} employs a feature-mapping network to convert input parameters into mel-spectrograms, which are then converted into the speech signals via a pre-trained neural vocoder. Subsequent work~\cite{hifiglot-baseline} uses a simplified feature-mapping model, enabling speaker-independent synthesis. 
However, as these models still follow a two-stage generation paradigm, explicit modeling of vocal tract characteristics and audio-level supervision are lacking, degrading parameter control precision and speech quality. 

To address these issues, this letter presents HiFi-Glot, an end-to-end neural speech synthesis system that uses source–filter modeling with differentiable all-pole filters.
HiFi-Glot implements the source-filter model by predicting the spectral envelope and glottal excitation, which will then be converted to the speech signal using an all-pole filter whose resonances are easy to inspect.
Both the spectral envelope and the glottal excitation are obtained from interpretable speech parameters using our proposed fully differentiable and computationally efficient methods. 
Experimental results show that our proposed HiFi-Glot achieves more precise control over formant frequencies and generates speech signals with higher perceptual quality and naturalness than baselines~\cite{praat, hifiglot-baseline}.

\section{Methodology}
\label{sec:method}

This section details the model architecture, training scheme, and differentiable all-pole filter designs of HiFi-Glot.

\subsection{Model Architecture}
\label{sec:modelarch}

As shown in Fig.~\ref{fig:model}, our proposed HiFi-Glot consists of a gated convolution neural network (GCNN)~\cite{hifiglot-baseline}-based feature mapper, an NSF-HiFiGAN decoder~\cite{diffsinger}, and four different discriminators. The feature mapping model consists of a group of cascaded, non-causal, gated CNN layers that convert the input speech parameters into all-pole filter parameters and a latent representation. The NSF-HiFiGAN then takes in the latent representation and the pitch contour to generate an excitation signal, which is subsequently passed through the predicted all-pole filter to produce the speech waveform.

To improve synthesis quality without damaging parameter controllability, we follow~\cite{cqtjournal} and add adversarial loss terms based on both time-domain and time-frequency-representation-based discriminators, specifically:
multi-period discriminator (MPD)~\cite{HiFiGAN}, multi-scale discriminator (MSD)~\cite{HiFiGAN}, multi-scale STFT discriminator (MS-STFTD)~\cite{dac}, and multi-scale sub-band CQT discriminator (MS-SB-CQTD)~\cite{cqt}.  

\subsection{All-Pole Synthesis Filter}

The all-pole synthesis filter is essential to the proposed system. Previous work on GAN-excited linear prediction (GELP)~\cite{gelp} enabled gradient propagation from the filtered output back to the excitation signal but was not differentiable with respect to the filter parameters. End-to-end LPCNet~\cite{e2elpcnet} addressed this by parameterizing reflection coefficients to ensure filter stability and gradient flow, but its efficiency is limited due to its autoregressive scheme. Here, we present a method that achieves both computational efficiency and full differentiability, enabling fast parallel synthesis on GPUs:

Treat a tensor as reflection coefficients $\bm k \in \mathbb{R}^{M \times P}$, where $M$ is the number of frames, $P$ is the filter order, and the batch dimension is omitted for simplicity. The direct-form linear predictor polynomial $\bm a \in \mathbb{R}^{M \times (P+1)}$ is then obtained using the forward Levinson recursion (which is differentiable). For parallel filtering, an FIR approximation of the synthesis filter response for each frame is obtained by:
\begin{equation}
\label{eq:all-pole-envelope}
\bm H(z) = \frac{\bm g}{\text{FFT}(\bm a, N) + \varepsilon} \in \mathbb{C} ^ {M \times N},
\end{equation}
where $N$ is the FFT length and $\bm g$ contains positive scalar gain terms for each frame. 
A time-domain excitation signal $\bm e(t) \in  \mathbb{R}^T$ is transformed to a complex spectrogram of matching size through a suitable STFT configuration, where:
\begin{equation}
    \bm E(z) = \text{STFT}(\bm e(t)) \in \mathbb{C}^{M \times N}.
\end{equation}
Filtering is then performed simply via complex multiplication in the STFT domain, obtaining:
\begin{equation}
    \bm X(z) = \bm H(z)  \bm E(z) \in \mathbb{C}^{M \times N}, 
\end{equation}
and a time-domain filtered signal is reconstructed via inverse FFT and overlap add, giving us the final result:
\begin{equation}
    \bm x(t) = \text{ISTFT}(\bm X(z)) \in \mathbb{R}^T.
\end{equation}

\subsection{Training Targets}

\begin{figure*}[ht!]
\centering
\includegraphics[width=0.90\linewidth]{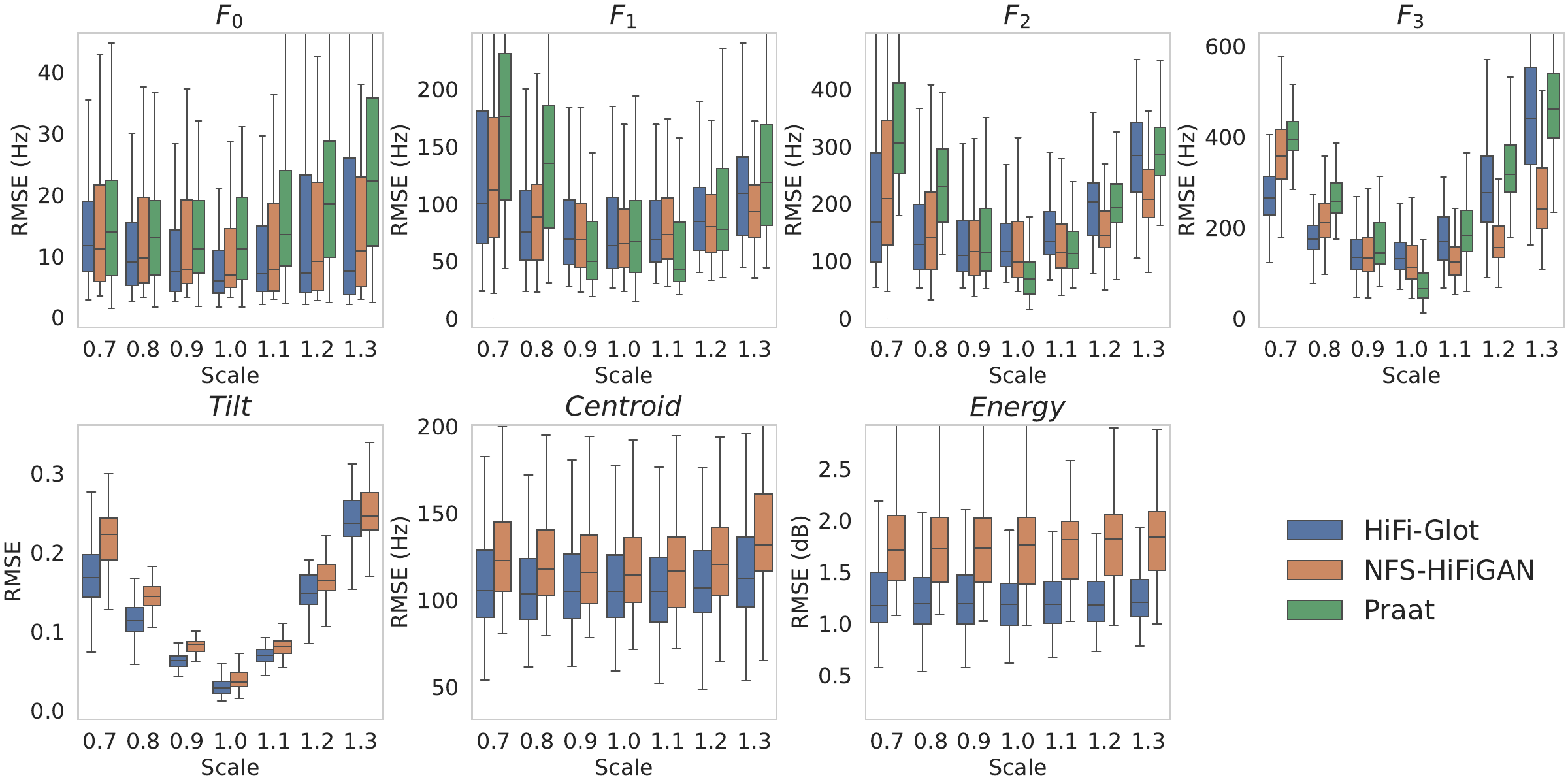}
\caption{Box plots of speech-parameter manipulation results. Each parameter is scaled over the range $[0.7, \dots, 1.3]$ and then re-synthesized using different systems. Box edges represent 1$^{\text{st}}$ and 3$^{\text{rd}}$ quartiles and whiskers extend to $1.5 \times \text{IQL}$. Note that all metrics are computed on voiced speech segments only, since we only manipulate speech parameter trajectories there. Additionally, Praat has no results for Tilt, Centroid, and Energy as it does not support manipulating these speech parameters.}
\label{fig:Man_objective}
\vspace{-8pt}
\end{figure*}

We adapted the training scheme from HiFi-GAN~\cite{HiFiGAN} to train the HiFi-Glot model, which is illustrated as follows:
\begin{equation}
\begin{split}
    &\mathcal{L}_\text{generator} = 15\mathcal{L}_\text{multi-mel}(x, \hat{x}) + \mathcal{L}_\text{envelope}(\bm H (z), \hat{\bm H}(z)) \\
    & + \sum\nolimits_{r=1}^{R}[\mathcal{L}_\text{adv}(G; D_{r}) + 2\mathcal{L}_\text{feat}(G; D_r)];
    \\
    &\mathcal{L}_\text{envelope} = \sum\nolimits_{m, n=1}^{M, N} || \log (|\bm H (z)|) - \log (| \hat{\bm H}(z)|) ||_1; \\
    &\mathcal{L}_\text{discriminator} = \sum\nolimits_{r=1}^{R}\mathcal{L}_\text{adv}(D_{r}; G); \\
\end{split}
\end{equation}
where $x$ and $\hat{x}$ are the ground-truth and predicted speech signal, $|\hat{\bm H}(z)|$ is the model output envelope spectrogram and $|\bm H (z)|$ is the ground truth LPC envelope spectrogram, $\mathcal{L}_\text{multi-mel}$ is the multi-scale mel-spectrogram loss adopted from DAC~\cite{dac}, $D_r$ is the $r^\text{th}$ discriminator, $G$ is the HiFi-Glot model, $\mathcal{L}_\text{adv}$ and $\mathcal{L}_\text{feat}$ are the adversarial losses and feature-matching loss, and $\mathcal{L}_\text{envelope}$ is the envelope spectrogram loss.

\section{Experiments}

\subsection{Experiment Setup}



\subsubsection{Datasets}

We used the same training set as in~\cite{afgen}, which contains 1664 hours of speech. We randomly selected 1000 utterances from HiFi-TTS2~\cite{hifitts2} for evaluation. 

\subsubsection{Baselines}

We used Praat~\cite{praat} for the DSP baseline system. For F0 manipulation, we used Parselmouth~\cite{parselmouth} for batch processing. For formant manipulation, we applied the following pipeline: 1) Estimate up to five formants, 2) adjust one or all of the first three formants in voiced speech segments, 3) combine the modified vocal tract envelope with the original excitation obtained via LPC inverse filtering, and 4) reimpose the original intensity trajectory on the modified speech.


For the neural-network baseline, we used the neural formant synthesis (NFS) model from~\cite{hifiglot-baseline}, denoted as NFS-HiFiGAN, and reimplemented it at a 44.1\,kHz sampling rate, with its decoder replaced by a mel-spectrogram pre-trained NSF-HiFiGAN model~\cite{diffsinger} (same as HiFi-Glot) for fair comparison.

\subsubsection{Configuration}

HiFi-Glot adapted the implementation of~\cite{hifiglot-baseline} to full-band operation. For the feature-mapping model, 8 GCNN layers were used, with residual and skip channels of 256 and a kernel size of 5. We used a 128-dimensional latent representation. We used 51 values to represent the predicted all-pole filter parameters, consisting of 50 log-area-ratios (LARs) and a gain factor. The envelope parameters were transformed to reflection coefficients using a tanh function to enforce
filter stability. We implemented the NSF-HiFiGAN decoder~\cite{diffsinger} with a modified multi-scale oscillator as in~\cite{neurodyne}, and used the code from Amphion~\cite{amphion} for the discriminators. We used periods of [2, 3, 5, 7, 11, 17, 23, 37] for the MPD; we computed 10 octaves and changed the hop sizes to [1024, 512, 512] for the MS-SB-CQTD, and left MSD, MS-STFTD, and other hyperparameters unmodified.

\subsubsection{Preprocessing}

We processed the training and evaluation datasets to 44.1\,kHz mono WAV files. $F_0$, $F_1$, $F_2$, $F_3$, and the voicing boundary were extracted using Praat~\cite{praat}, the spectral tilt was extracted as the predictor coefficient of the first-order all-pole polynomial, the spectral centroid was extracted as the average of frequency bin values weighted by their corresponding frequency value, and the energy was a log-scale value of the frame energy. All features were extracted using a Hann window with a length of 2048 and a hop size of 512. $F_0$, $F_1$, $F_2$, $F_3$, and spectral tilt values were linearly interpolated in unvoiced regions. Formant contours were further smoothed using a median filter with a kernel size of 7. All features were then min-max normalized to [$-$1, 1].

\subsubsection{Training}

All the models were trained using the AdamW optimizer with $\beta _ {1} = 0.8$ and $\beta _ {2} = 0.99$, a learning rate of 2e$-$5, and an exponential decay scheduler with a factor $\gamma = 0.999996$. All the experiments were done on an H200 GPU with a batch size of 16 and a 1\,s segment length for 1M steps.

\begin{figure*}[ht!]
\centering
\includegraphics[width=\linewidth]{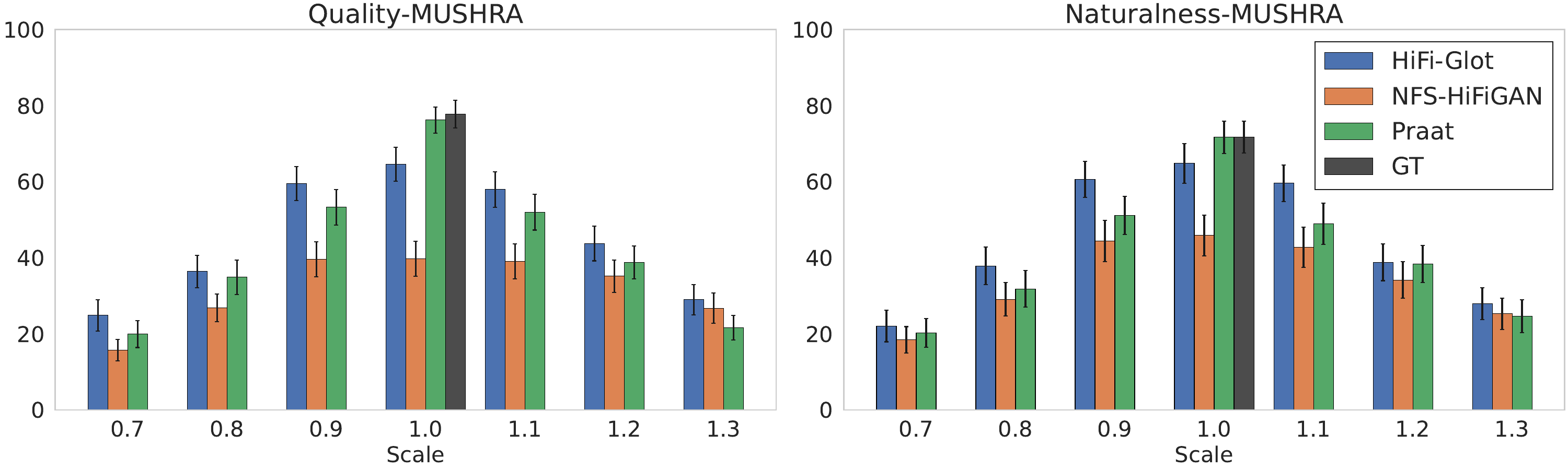}
\caption{Bar plots of the MUSHRA-like listening test results with 95\% confidence intervals (CIs). All formant parameters are scaled over the range $[0.7, \dots, 1.3]$ and then re-synthesized using different systems.}
\label{fig:Man_subjective}
\end{figure*}

\subsection{Evaluation Metrics}

We computed the root-mean-square error (RMSE) between the estimated feature values of the re-synthesized and ground-truth speech for objective evaluation. 
For subjective evaluation, we employed a MUSHRA-like test~\cite{mushra} comprising 35 distinct test trials. In each trial, synthesized samples from all systems for a given utterance were presented on the same screen to allow for side-by-side comparison.
Unlike the standard MUSHRA test, a labeled ground-truth reference and a 16\,kHz-resampled hidden anchor were provided only when a ground-truth recording was available; otherwise, no reference was presented. 
Listeners were asked to provide scores (denoted as Quality- and Naturalness-MUSHRA) ranging from 0 to 100. 
To ensure data reliability, we integrated 5 attention-check trials in which listeners were explicitly instructed by the audio to grade a sample within a specific score range; failure at any check resulted in exclusion.
We recruited 20 participants via Prolific, screening for native English speakers with an approval rate of at least 95\%. Participants who passed all attention-check trials received 12 EUR as compensation.


\subsection{Effectiveness on Speech Parameter Manipulation}

Fig.~\ref{fig:Man_objective} illustrates the manipulation errors for different speech parameters across various scaling factors. For Tilt, Centroid, and Energy, HiFi-Glot consistently outperformed the baseline model, with lower median errors across all scaling factors. Regarding $F_0$, HiFi-Glot outperformed the baselines across most scaling factors. The only exception is at the 0.7 scale, where it performed comparably to NFS-HiFiGAN, though it remained superior to Praat. For $F_1$, $F_2$, and $F_3$, HiFi-Glot demonstrated clear advantages when the scaling factors were below 1.0 (downward scaling). 
Meanwhile, when the scaling factors were greater than or equal to 1.0 (upward scaling), HiFi-Glot performed on par with the NFS-HiFiGAN baseline while consistently outperforming Praat. Overall, these results validated the effectiveness of the proposed HiFi-Glot model, highlighting its exceptional accuracy in downward scaling and competitive performance in upward scaling.
Note that the Praat baseline benefits from access to the original excitation signal and the LPC envelope, and it simply performs LPC copy-synthesis at a unit-scale (1.0) modification. In contrast, neural methods need to generate signals directly from the speech parameters; despite this, our model remains competitive.

\subsection{Effectiveness on Vocal Tract Length Modification}

We evaluate the perceptual quality and naturalness of the proposed method in a vocal tract length modification scenario using a MUSHRA-like listening test. We simulate changes in vocal tract length by scaling all formant parameters ($F_1$, $F_2$, and $F_3$) simultaneously by factors in the range $[0.7, \dots, 1.3]$. The subjective evaluation results are shown in Fig.~\ref{fig:Man_subjective}. 
Note that the relatively low absolute scores are expected, as globally scaling formant parameters inherently results in speech that sounds less natural compared to the original recordings.

While Praat achieves the highest scores at the unit scale (1.0, i.e., no manipulation) by essentially applying LPC copy-synthesis, its performance degrades rapidly as the scaling factor deviates from unity.
For all these scaling factors, speech manipulation using HiFi-Glot has superior performance and robustness in terms of both quality and naturalness, compared to both Praat and NFS-HiFiGAN. This confirms that our proposed HiFi-Glot offers a better trade-off between high-fidelity reconstruction and flexible formant control. 

\section{Conclusion}

This letter presents HiFi-Glot, an end-to-end neural formant synthesis system that combines a source-filter architecture with resonant all-pole filters. Through a fully differentiable filter design, HiFi-Glot enables accurate, interpretable control of key speech parameters, including formants, spectral tilt, centroid, energy, and pitch. Experimental results illustrate that our proposed HiFi-Glot can generate speech of higher quality and naturalness, with superior manipulation accuracy, compared with baseline systems, confirming its effectiveness. 

\newpage



\bibliographystyle{IEEEtran}
\bibliography{ref}

\end{document}